\documentclass[aps,prd,amsmath,twocolumn,amssymbaps,showpacs]{revtex4-2}
\usepackage{graphicx}  
\usepackage{dcolumn}   
\usepackage{bm}        
\usepackage{amssymb}   
\usepackage{amsmath}   
\usepackage{bbm}
\usepackage{draftcopy}
\usepackage{relsize} 
\usepackage{xfrac}    
\usepackage{slashed}

\usepackage{color}
\definecolor{dgreen}{cmyk}{1.,0.,1.,0.2}        
\definecolor{orange}{cmyk}{0.,0.353,1.,0.}    

\usepackage[bookmarks]{hyperref}

\newcommand{\di}{{\rm d}}

\newcommand{\be}{\begin{equation}}
\newcommand{\ee}{\end{equation}}                                                                               
\newcommand{\bea}{\begin{eqnarray}}
\newcommand{\eea}{\end{eqnarray}}

\begin{document}
\title{Extended Nambu--Jona-Lasinio model for quark and nuclear matters}

\author{Gaoqing Cao}
\affiliation{School of Physics and Astronomy, Sun Yat-sen University, Zhuhai 519088, China}

\date{\today}

\begin{abstract}
{In this work, we extend the two-flavor Nambu--Jona-Lasinio model to one capable of exploring quark and nuclear matter consistently. With an extra term standing for quark-nucleon interactions, nucleons could automatically emerge as color-singlet three-quark entities by following a process similar to mesons. Besides the quark part in mean field approximation, both mesons and nucleons could contribute to the thermodynamic potential thus possibly give rise to quarkyonic matter beyond mean field. In the study, two kinds of "confining" couplings are adopted for the new interaction term and two different quark masses are considered for comparison. It turns out that only confined nuclear matter or deconfined quark matter is possible for all the cases at zero temperature, thus quarkyonic matter is not favored at all. Even more strictly, only the case with stronger confinement effect and a smaller quark mass admits a physical first-order phase transition from nuclear matter to quark matter around twice saturation density.} 
\end{abstract}


\maketitle

\section{Introduction}
In 1961, Y. Nambu and G. Jona-Lasinio developed a dynamical model of nucleons to explain the origin of nucleon mass based on an analogy with superconductivity~\cite{Nambu:1961tp,Nambu:1961fr}. They had shown that a significant dynamical mass can be generated through chiral condensation, with which the approximate chiral symmetry of the quantum chromodynamics (QCD) vacuum is broken~\cite{Nambu:1961tp}. The model was then named after the authors as "Nambu--Jona-Lasinio (NJL) model". However, with the discovery of quarks as more fundamental blocks of hadrons in 1964~\cite{Gell-Mann:1964ewy,Zweig:1964ruk}, the NJL model had been modified to one with only quark degrees of freedom~\cite{Klevansky:1992qe,Hatsuda:1994pi}. Though the property of confinement is lost, such a model could give a more physical picture on the quark-antiquark structures of mesons and is successful in reproducing the well-known Gell-Mann–Oakes–Renner relations of QCD~\cite{Klevansky:1992qe}. Therefore, in high energy nuclear physics, the quark NJL model is more welcome and is frequently adopted to study QCD phase transition, such as chiral symmetry breaking ($\chi$SB) and restoration ($\chi$SR)~\cite{Klevansky:1992qe,Hatsuda:1994pi}.

In mean field approximation, the quark NJL model predicts that the approximate chiral symmetry gets restored through a crossover at finite temperature ($T$) and zero baryon chemical potential ($\mu_{\rm B}$) but a first-order transition at zero $T$ and finite $\mu_{\rm B}$~\cite{Klevansky:1992qe,Hatsuda:1994pi}. So, a critical point is expected at medium $T$ and $\mu_{\rm B}$, the values of which are now a very hot topic in both experimental and theoretical studies~\cite{Luo:2017faz,Luo:2022mtp}. Other physical parameters can also be involved in the study of QCD phase transition, such as isospin chemical potential~\cite{He:2005nk,Sun:2007fc,Cao:2016ats} and magnetic field~\cite{Miransky:2015ava,Andersen:2014xxa,Cao:2021rwx}. Here, we only focus on $T-\mu_{\rm B}$ phase digram. Since $u/d$ quark mass is greatly reduced at large $\mu_{\rm B}$, a Fermi sea of quarks is expected at zero $T$~\cite{Klevansky:1992qe,Hatsuda:1994pi}. However, if confinement resumes there as in vacuum and nuclear matter, the effective excitations must be the color-singlet baryons, which means that quarks must triply pair to baryons on the Fermi surface. Such a quark-baryon coexistent state was first proposed by following a large $N_{\rm c}$ argument and named as "quarkyonic matter"~\cite{McLerran:2007qj}. Recently, such an exotic matter is a very hot topic in nuclear physics~\cite{McLerran:2018hbz,Cao:2020byn,Pisarski:2021aoz,Sen:2020qcd,Cao:2022inx,Koch:2022act,Fujimoto:2023mzy}.

So in order to study cold, dense QCD matter more realistically, how could nucleon consistently emerge from the quark NJL model? Previously, people extended the quark NJL model by including a four-quark interaction in the most attractive diquark channels~\cite{Alford:2007xm} in a similar way as the mesonic channels. Then, the properties of nucleon can be explored by following the Faddeev equations, where nucleon is considered as a bound state of diquark and quark~\cite{Buck:1992wz,Ishii:1995bu}. In such a model, if a finite $\mu_{\rm B}$ is introduced, we might find a coexistent system with quarks, diquarks and nucleons, inconsistent with the picture of quarkyonic matter~\cite{McLerran:2007qj}. To avoid that, we extend the quark NJL model to include a realistic six-quark interaction in nucleonic channel which maintains the basic symmetries of QCD, and then nucleons could emerge without involving diquarks. Beyond mean field approximation, the QCD system could be a coexistent state of quarks, mesons and nucleons in principle. Hence, the possibility of quarkyonic matter and transitions among nuclear, quark and quarkyonic matters can be consistently explored within such a unified model. {Note that there was previously a similar study in quark NJL model where both diquark and baryon channels are involved~ \cite{Wang:2010iu}, but diquarks are not necessarily to be stable compared to Faddeev method.} The method developed here for three-fermion excitation could also be applied to other branches of physics, such as cold atoms and condensed matter.

The paper is organized as follows. In Sec.~\ref{model}, the extended NJL model is discussed and overall formalism is developed for the study of chiral symmetry breaking and 
restoration. In Sec.~\ref{calculation}, by choosing two different quark masses and two different couplings for the six-quark interaction term, numerical calculations are carried out mainly 
to explore the features of chiral symmetry and system constituents at zero $T$ and finite $\mu_{\rm B}$. Finally, a brief summary would be given in Sec.~\ref{summary}.

\section{NJL model with nucleonic interaction channel}\label{model}

\subsection{The extended Lagrangian}
The Lagrangian density of two-flavor NJL model with quark degrees of freedom was given by~\cite{Klevansky:1992qe,Hatsuda:1994pi}
\bea
{\cal L}_0&=&\bar\psi\left(\slashed{\partial}-m_0\right)\psi+G_2\left[\left(\bar\psi\psi\right)^2+\left(\bar\psi i\gamma_5\boldsymbol\tau\psi\right)^2\right],\label{L0}
\eea
where $\psi(x)=(u(x),d(x))^T$ is the two-flavor quark field and ${\boldsymbol\tau}\equiv(\tau_1,\tau_2,\tau_3)$ is the Pauli matrix vector in isospin space. In this model, the $SU_A(2)$chiral symmetry is slightly violated by the small current quark mass $m_0$ but would be significantly broken with chiral condensation due to the strong coupling constant $G_2$.

Now, we are going to introduce a six-quark interaction term that accounts for three quarks interacting through nucleon channels. Then first of all, how can we present a nucleon in term of three-quark current? According to the discussions in the QCD sum rules~\cite{Ioffe:1981kw,Ioffe:1982ce}, the most suitable derivative-free quark current for a color singlet nucleon is the one with an axial vector diquark coupled to a quark, that is, 
\bea 
p &\propto& \epsilon_{abc}(u^{aT}C\gamma_\mu u^b)\gamma_5\gamma^\mu d^c,\\ 
n&\propto& \epsilon_{abc}(d^{aT}C\gamma_\mu d^b)\gamma_5\gamma^\mu u^c
\eea
with $C=i\gamma_2\gamma_0$. Here, $\epsilon_{abc}$ is the Levi-Civita symbol in color space that makes sure a color singlet of nucleon. Later, it was found that each form can be separated into two independent terms~\cite{Jin:1993up}: the ones with scalar and pseudoscaler diquarks coupled to a quark, that is, 
\bea 
p &\propto& \epsilon_{abc}(u^{aT}C\gamma_5d^b)u^c-\epsilon_{abc}(u^{aT}C d^b)\gamma_5u^c,\\ 
n&\propto& \epsilon_{abc}(u^{aT}C\gamma_5d^b)d^c-\epsilon_{abc}(u^{aT}C d^b)\gamma_5d^c.
\eea
With the developments of lattice QCD simulations, we realized that the scalar diquark channel is the most attractive one, so it is reasonable and convenient to present the two-flavor nucleon field compactly as 
\bea
N\equiv\left(\begin{array}{c}
	p \\ n
\end{array}\right)\propto \psi^{\rm c}\left(\psi^{\rm T}C\tau_2\epsilon_{\rm c}\gamma_5\psi\right).
\eea 
Here, $\epsilon_{\rm c}$ are the antisymmetric matrices in color space, that is, $(\epsilon_{\rm c})_{ab}=\epsilon_{abc}$ with indices $a,b,c=(r,g,b)$ and $\epsilon_{abc}$ the Levi-Civita symbol.

Next, how can we present the six-quark interaction term? As we can check, the effective four quark interaction terms in Eq.\eqref{L0}, corresponding to $\sigma$ and pion channels, completely respect chiral symmetry and Lorentz invariance of the basic QCD theory. To keep these symmetries, the simplest six-quark interaction term can be given as
\bea
-\left(\bar\psi\tau_2\epsilon_{\rm c}\gamma_5\psi_C\right)\bar\psi^{\rm c}\ {\tilde{G}_3} i\slashed{\partial}\ \psi^{\rm c'}\left(\bar\psi_C\tau_2\epsilon_{\rm c'}\gamma_5\psi\right),
\eea
where $\psi_C=C\bar\psi^T$ and $\bar\psi_C=\psi^TC$ are charge-conjugate spinors, and the coupling constant $\tilde{G}_3$ could be $\partial^2$ dependent.
Finally, at finite temperature and baryon chemical potential, the Lagrangian of the extended NJL model can be given in four momentum space as
\bea
{\cal L}&=&\bar\psi\left(\slashed{q}+\gamma^0{\mu_{\rm B}\over3}-m_0\right)\psi+G_2\left[\left(\bar\psi\psi\right)^2+\left(\bar\psi i\gamma_5\boldsymbol\tau\psi\right)^2\right]\nonumber\\
&&-{\tilde{G}_3}\left(\bar\psi\tau_2\epsilon_{\rm c}\gamma_5\psi_C\right)\bar\psi^{\rm c}\slashed{\tilde{P}}\psi^{\rm c'}\left(\bar\psi_C\tau_2\epsilon_{\rm c'}\gamma_5\psi\right),\label{Lq}
\eea
where $\tilde{P}_\nu\equiv P_\nu+\mu_{\rm B}\,\delta_{\nu 0}$ with $P=q_1+q_2+q_3$. As can be seen, the coupling is effectively ${\tilde{G}_3}\tilde{P}$ for the six-quark interaction term and would vanish in the limit $|\tilde{P}|\rightarrow0$ if ${\tilde{G}_3}$ is a constant. To respect the confinement feature of QCD, $\lim_{|\tilde{P}|\rightarrow0}{\tilde{G}_3}|\tilde{P}|=\infty$ must be assumed, so we will take ${\tilde{G}_3}={G}_3/|\tilde{P}^2|^{d}\ (d=1,3/2)$ with $G_3$ a constant for example in the study. Though confinement seems not so important for the spectra of mesons~\cite{Klevansky:1992qe,Hatsuda:1994pi}, it is important in order to obtain reasonable propagators and spectra of nucleons.

\subsection{Mean field approximation}
In the standard processes, chiral condensation is usually studied in mean field approximation and the properties of mesons are then explored in random phase approximation (RPA)~\cite{Klevansky:1992qe,Hatsuda:1994pi}. The mesonic degrees of freedom can be introduced through Hubbard-Stratonavich transformation and the Lagrangian Eq.\eqref{Lq} becomes
\bea
{\cal L}&=&\bar\psi\left(\slashed{q}+\gamma^0{\mu_{\rm B}\over3}-m_0-\sigma-i\gamma_5\boldsymbol{\tau\cdot\pi}\right)\psi-{\sigma^2+\boldsymbol\pi^2\over4G_2}\nonumber\\
&&-{\tilde{G}_3}\left(\bar\psi\tau_2\epsilon_{\rm c}\gamma_5\psi_C\right)\bar\psi^{\rm c}\slashed{\tilde{P}}\psi^{\rm c'}\left(\bar\psi_C\tau_2\epsilon_{\rm c'}\gamma_5\psi\right)
\eea
with $\sigma\equiv-2G_2\bar\psi\psi$ and $\boldsymbol\pi\equiv-2G_2\bar\psi i\gamma_5\boldsymbol\tau\psi$. In the mean field approximation with $\langle\sigma\rangle=m-m_0\neq0$, the thermodynamic potential can be easily evaluated by integrating out quark degrees of freedom as
\bea
\Omega_0&=&{(m-m_0)^2\over 4G_2}-2N_{\rm f}N_{\rm c}\int^\Lambda{\di^3k\over(2\pi)^3}E_{\bf k}-2TN_{\rm f}N_{\rm c}\nonumber\\
&&\times\sum_{t=\pm}\int{\di^3k\over(2\pi)^3}\ln\left[1+e^{-{1\over T}(E_{\bf k}+t\,{\mu_{\rm B}\over3})}\right].\label{omg0}
\eea
Here, the quark energy $E_{\bf k}=({\bf k}^2+m^2)^{1/2}$, $\Lambda$ is the three-momentum cutoff, $N_{\rm f}$ and $N_{\rm c}$ are flavor and color degrees of freedom of quarks, respectively.
At this level, the dynamical mass $m$ is then determined according to the gap equation given by $\partial\Omega_0/\partial m=0$.
\subsection{Beyond Mean field approximation}
Based on the mean field approximation, the propagator of a quark with color $c$ is 
\bea
S_{\rm q}^{\rm c}&\equiv&i/(\slashed{q}+\gamma^0{\mu_{\rm B}/3}-m).\label{qp}
\eea
Then, by taking the bare propagators of mesons to be $-2iG$, their full inverse propagators can be evaluated through RPA approximation as~\cite{Klevansky:1992qe}
\bea
S_{\sigma/\boldsymbol\pi}^{-1}(p)={i\over2G}-{\rm Tr}\ S_{\rm q}^{\rm c}(q)\Gamma_{\sigma/\boldsymbol\pi}S_{\rm q}^{\rm c}(q-p)\Gamma_{\sigma/\boldsymbol\pi}
\eea
with $\Gamma_{\sigma}=1$, $\Gamma_{\boldsymbol\pi}=i\gamma_5\boldsymbol\tau$, and the trace ${\rm Tr}$ over Dirac, flavor, color, and momentum spaces. After substituting Eq.\eqref{qp} into this equation, the pole mass of pseudo-Goldstone pions can be evaluated by requiring $S_{\boldsymbol\pi}^{-1}(p_0=m_\pi,{\bf p=0})=0$, and we have
\bea
0&=&{1\over2G}-2N_{\rm f}N_{\rm c}\int^\Lambda{\di^3k\over(2\pi)^3}{E_{\bf k}\over E_{\bf k}^2-m_\pi^2/4}+2N_{\rm f}N_{\rm c}\nonumber\\
&&\times\sum_{t=\pm}\int{\di^3k\over(2\pi)^3}{E_{\bf k}\over E_{\bf k}^2-m_\pi^2/4}{1\over1+e^{{1\over T}(E_{\bf k}+t\,{\mu_{\rm B}\over3})}}.\label{pimass}
\eea
Then, beyond mean field approximation, the contribution of pions to the thermodynamic potential can be evaluated in pole approximation as
\bea
\Omega_{\rm \pi}=3T\int{\di^3k\over(2\pi)^3}\ln\left(1-e^{-{E_{\bf k}^{\rm \pi}\over T}}\right).
\eea
with $E_{\bf k}^\pi=({\bf k}^2+m_\pi^2)^{1/2}$. In the case of high isospin density, charged pions are actually collective excitations of quark-antiquark pairs around the Fermi surface of (anti-)quarks and also contribute to isospin density~\cite{Sun:2007fc,Cao:2016ats}.

\begin{figure}[!htb]
	\begin{center}
	\includegraphics[width=8cm]{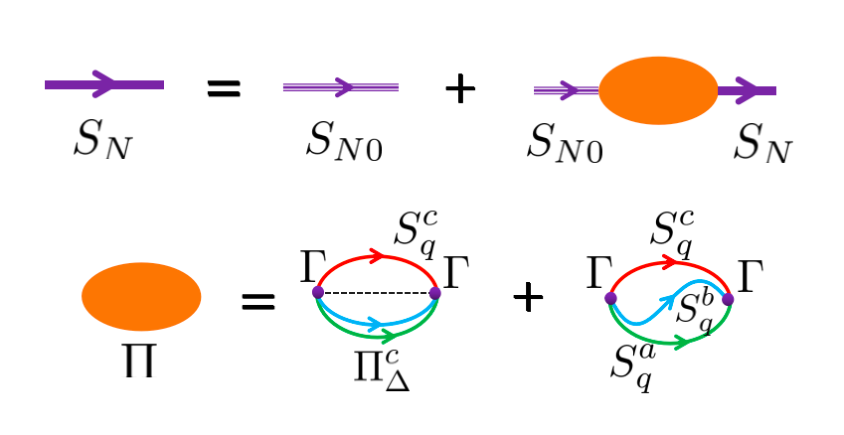}
		\caption{The diagrammatic representation of the random phase approximation for the full nucleon propagator $S_{\rm N}$. }\label{SN_RPA}
	\end{center}
\end{figure}

By following the same logic as the Hubbard-Stratonavich transformation, we introduce the nucleon degree of freedom as $N\equiv \sqrt{{\tilde{G}_3}}\psi^{\rm c}\left(\psi^{\rm T}C\tau_2\epsilon_{\rm c}\gamma_5\psi\right)$, and the six-quark interaction term in Eq.\eqref{Lq}  can be equivalently rewritten as
\bea
{\cal L}_6&=&\bar{N}\ i\,S_{\rm N0}^{-1}\ N-\bar{N}\ i\,\Gamma\ \psi^{\rm c}\left(\bar\psi_C\tau_2\epsilon_{\rm c}\gamma_5\psi\right)\nonumber\\
&&-\left(\bar\psi\tau_2\epsilon_{\rm c}\gamma_5\psi_C\right)\bar\psi^{\rm c}\ i\,\Gamma\ N.
\eea
Here, the bare nucleon propagator $S_{\rm N0}=i/\slashed{\tilde{P}}$ is in a massless form, and we have defined the quark-nucleon interaction vertex as $\Gamma(P,\mu_B)\equiv \sqrt{{\tilde{G}_3}}S_{\rm N0}^{-1}$. To explore nucleonic properties, the full propagator of nucleon has to be evaluated by taking into account the interaction with quarks. Inspired from the cases with mesons, a RPA-like scheme can be adopted to derive the full propagator, $S_{\rm N}\equiv \langle{\cal T}N\bar{N}\rangle$ (${\cal T}$ is the time-ordering operator), see the diagrammatic representation in Fig.\ref{SN_RPA}.
Regardless of the vertices, the first and second diagrams for the self-energy actually correspond to the free and one-quark exchange diagrams in Fadeev equations~\cite{Buck:1992wz}. However, with the vertices, three quark propagators are directly involved in the evaluation of self-energy but no diquark propagator is necessary to bridge the emergence of nucleons from quarks, contrary to the Fadeev scheme~\cite{Buck:1992wz,Ishii:1995bu,Eichmann:2016yit}. 

Translate Fig.\ref{SN_RPA} to mathematical expressions, we have
\bea
&&S_{\rm N}=S_{\rm N0}(1+\Gamma\, {\Pi}\,\Gamma S_{\rm N})=S_{\rm N0}+\bar{\Pi} S_{\rm N0}^{-1}S_{\rm N}, \label{SN0}\\
&&\bar{\Pi}(P) \equiv\Pi_{\rm q\Delta}(P)+\Pi_{\rm qqq}(P),\\
&&\Pi_{\rm q\Delta}(P)={\tilde{G}_3}\int{\di^4 l\over (2\pi)^4}S_{\rm q}^{\rm c}(P-l)\Pi_\Delta^{\rm c}(l),\label{PiqD}\\
&&\Pi_{\rm qqq}(P)={\tilde{G}_3}\int{\di^4 l\over (2\pi)^4}{\di^4 k\over (2\pi)^4}\tilde\Pi_{\rm qqq}(P,l,k),\label{Piqqq0}
\eea
where the self-energies of diquark and three-quark irreducible correlation are, respectively,
\bea
&&\Pi_\Delta^{\rm c}(l)\equiv\int{\di^4 k\over (2\pi)^4}{\rm Tr} \ \tau_2\epsilon_{\rm c}C\gamma_5 S_{\rm q}(l-k)\tau_2\epsilon_{\rm c}\gamma_5CS_{\rm q}^{\rm T}(k),\label{PiD}\\
&&\tilde\Pi_{\rm qqq}\equiv3 [S_{\rm q}(P-l)(\tau_2\gamma_5C)S_{\rm q}^{\rm T}(l-k)(C\tau_2\gamma_5)^{\rm T}S_{\rm q}(k)\nonumber\\
&&\ \ \ \ \ \ \ \ \ \ \ \ \ \ \ +S_{\rm q}(P-l)(\tau_2\gamma_5C)^TS_{\rm q}^{\rm T}(l-k)(C\tau_2\gamma_5)S_{\rm q}(k)]\nonumber\\
&&\ \ \ \ \ \ \ =-6S_{\rm q}(P-l)S_{\rm q}(l-k)S_{\rm q}(k).\label{Piqqq}
\eea
Note that in Eqs.\eqref{PiD} and \eqref{Piqqq}, the color indices of quark propagators are suppressed as they are identical in color space. Again, though the self-energy of diquark shows up in the RPA study, diquarks are never independent degrees of freedom in our formalism, rather, must show together with the third quark. With the full self-energy $\bar{\Pi}(P)$ evaluated, the full nucleon propagator can then be explicitly derived from Eq.\eqref{SN0} as
\bea
S_{\rm N}=i\,S_{\rm N0}^{-1}\left[i\,S_{\rm N0}^{-1}+i\,\bar{\Pi}(P)\tilde{P}^2\right]^{-1}S_{\rm N0}.\label{SN}
\eea
The regularized integrals in $\bar{\Pi}(P)$ are usually finite, so if we choose ${\tilde{G}_3}\equiv{G}_3|\tilde{P}^2|^d$ with $d>-1/2$, we would find $\lim_{|\tilde{P}^2|\rightarrow 0}S_{\rm N}=S_{\rm N0}$ and zero mass becomes a trivial pole of the propagator. That is of course not physical as the true nucleon mass is large in the vacuum -- this is another reason why we choose $d=1,3/2$ in the study.

The self-energies $\Pi_{\rm q\Delta}(P)$ and ${\Pi}_{\rm qqq}(P)$ are explicitly evaluated in Appendix \ref{APiqd} and Appendix \ref{APiqqq}, respectively, by substituting the quark propagator \eqref{qp}. To calculate nucleon pole mass, it is conventional to set the three-momentum ${\bf P}$ to be ${\bf 0}$ in the propagator. In that case, only the time components $\Pi_{\rm q\Delta /qqq}^{\rm 0}(\tilde{P}_0)$ and scalar components $\Pi_{\rm q\Delta /qqq}^{\rm s}(\tilde{P}_0)$ survive, refer to Eqs. \eqref{EPiqd} and \eqref{EPiqqq}; and we have $\bar{\Pi}(\tilde{P}_0)=i\,[\gamma^0\Pi^0(\tilde{P}_0)+\Pi^{\rm s}(\tilde{P}_0)]$ with $\Pi^{\rm0/s}(\tilde{P}_0)=\Pi_{\rm q\Delta}^{\rm0/s}(\tilde{P}_0)+\Pi_{\rm qqq}^{\rm0/s}(\tilde{P}_0)$. According to Eqs. \eqref{Piqds} and \eqref{Piqqqs}, $\Pi^{\rm s}(\tilde{P}_0)\propto m$, so the pattern of chiral symmetry is consistently controlled by the dynamical quark mass $m$ even for the nucleon sector.

Due to the commutativity between time  and scalar components, the nucleon propagator Eq.\eqref{SN} can be reduced to
\bea
S_{\rm N}&=&{i\over\gamma^0[\tilde{P}_0-\bar{\Pi}_0(\tilde{P}_0)\tilde{P}_0^2]-\bar{\Pi}_{\rm s}(\tilde{P}_0)\tilde{P}_0^2}.
\eea
So, the pole mass of nucleons can be obtained by solving the following equation self-consistently:
\bea
[m_{\rm N}-\bar{\Pi}_0(m_{\rm N})m_{\rm N}^2]^2-\bar{\Pi}_{\rm s}^2(m_{\rm N})m_{\rm N}^4=0,\label{Nmass}
\eea
and their contribution to the thermodynamic potential can be evaluated in pole approximation as~\cite{Zhuang:1994dw}
\bea
\Omega_{\rm N}=-2TN_{\rm f}\sum_{t=\pm}\int{\di^3k\over(2\pi)^3}\ln\left[1+e^{-{1\over T}(E_{\bf k}^{\rm N}+t\,{\mu_{\rm B}})}\right]
\eea
with $E_{\bf k}^{\rm N}=({\bf k}^2+m_{\rm N}^2)^{1/2}$. In this way, nucleons are naturally the collective excitations of three quarks around the Fermi surface of quarks by following a similar picture as pions at large isospin density. Thus, quarkyonic matter could be systematically realized if considerable nucleons are found to coexist with a large density of quarks in the $\chi$SR phase.

In total, the beyond mean field thermodynamic potential is 
\bea
\Omega\equiv \Omega_0+\Omega_\pi+\Omega_{\rm N}.\label{omg}
\eea 
In principle, beyond-mean-field dynamical quark mass can be self-consistently obtained by finding the global minimum of $\Omega$ with pion and nucleon masses solved from Eqs.~\eqref{pimass} and \eqref{Nmass}. At zero temperature, the contributions of quarks and nucleons to the baryon density can be evaluated according to $n_{\rm B}=-\partial \Omega/\partial \mu_{\rm B}=n_{\rm Bq}+n_{\rm BN}$ as
\bea
\!\!\!\!\!\!n_{\rm Bq}\equiv-{\partial \Omega_0\over\partial \mu_{\rm B}}={N_{\rm f}\over3\pi^2}p_{\rm Fq}^3,\
n_{\rm BN}\equiv-{\partial \Omega_{\rm N}\over\partial \mu_{\rm B}}={N_{\rm f}\over3\pi^2}p_{\rm FN}^3
\eea
with the Fermi momenta
\bea
p_{\rm Fq}=\sqrt{(\mu_{\rm B}/N_{\rm c})^2-m_{\rm q}^2}, \
p_{\rm FN}=\sqrt{\mu_{\rm B}^2-m_{\rm N}^2}.
\eea
For $m_{\rm N}= N_{\rm c}m_{\rm q}$, $p_{\rm FN}=N_{\rm c}\,p_{\rm Fq}$ and thus $n_{\rm BN}=N_{\rm c}^3n_{\rm Bq}$. It indicates that nucleons are more efficient to establish high baryon density when $m_{\rm N}\sim N_{\rm c}m_{\rm q}$ as is the case for nuclear matter.
\section{Numerical calculation}\label{calculation}
By following the scheme of Ref.~\cite{Zhuang:1994dw}, the model parameters could be fixed as
$$m_0=5.29~{\rm MeV}, \ G_2=4.9316~{\rm GeV}^{-2},\ \Lambda=0.65333~{\rm GeV}$$
 by fitting to the pion mass $m_\pi=0.138~{\rm GeV}$, the pion decay constant $f_\pi=0.093~{\rm GeV}$ and the chiral condensates $\langle\bar{\rm u}{\rm u}\rangle=\langle\bar{\rm d}{\rm d}\rangle=-(0.25~{\rm GeV})^3$ in the vacuum. The dynamical quark mass in vacuum is then $m_{\rm q}=0.3135~{\rm GeV}$, a bit larger than $m_{\rm N}/3$ with $m_{\rm N}=0.939~{\rm GeV}$ the nucleon vacuum mass. Note that the condition $m_{\rm q}>m_{\rm N}/3$ is necessary to guarantee that quarks appear later than nucleons with the increasing of  $\mu_{\rm B}$  in the QCD system. That is consistent with our common sense that quarks are confined in nucleons in the vacuum. Then, the six-quark coupling constant can be fixed as ${G_3 m_{\rm N}^{2-2d}}=2284~{\rm GeV}^{-4}$ by fitting to $m_{\rm N}=0.939~{\rm GeV}$. In the following, we would refer to such a set of parameters as "small quark vacuum mass (SQVM)". For comparison, we modify the chiral condensate a little to $\langle\bar{\rm u}{\rm u}\rangle=\langle\bar{\rm d}{\rm d}\rangle=-(0.249~{\rm GeV})^3$ and the model parameters alter correspondingly to
 \bea
&& m_0=5.37~{\rm MeV},\ \ \ \ \ \ \ \Lambda=0.6455~{\rm GeV},\nonumber\\
&&\ G_2=5.112~{\rm GeV}^{-2}, \ G_3m_{\rm N}^{2-2d}=2375~{\rm GeV}^{-4}.
 \eea
 As $m_{\rm q}=0.320~{\rm GeV}$ in vacuum, we refer to such a set of parameters as "large quark vacuum mass (LQVM)" in the following.
 
 \subsection{Small quark vacuum mass}
\begin{figure}[!htb]
	\begin{center}
	\includegraphics[width=8.5cm]{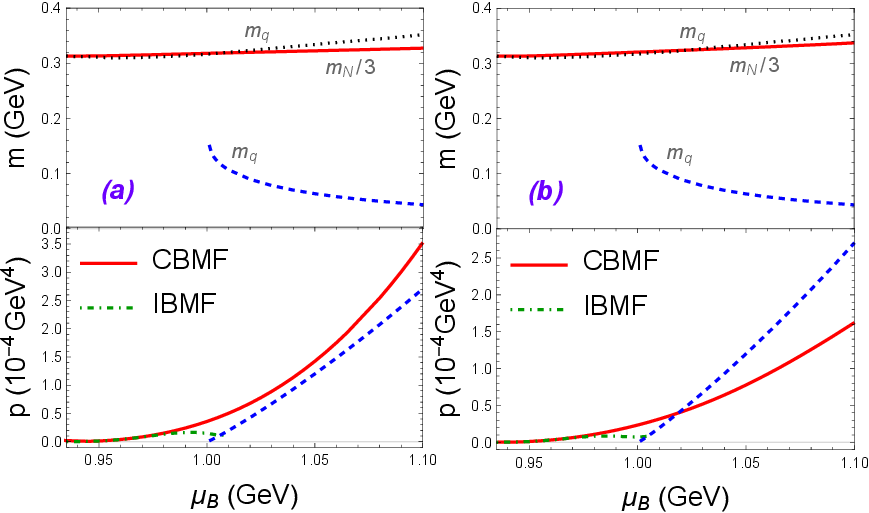}
		\caption{The SQVM case with $(a)\ d=1$ and $(b)\ d=3/2$ for ${\tilde{G}_3}$. Upper panels: the quark mass $m_{\rm q}$ (black dotted) and nucleon mass $m_{\rm N}$ (red solid) as functions of the baryon chemical potential $\mu_{\rm B}$  for $\chi$SB phase in the CBMF approximation, and the quark mass $m_{\rm q}$ in $\chi$SR phase (blue dashed). Lower panels: the pressures for $\chi$SB phase in the CBMF (red solid) and IBMF (green dashdotted) approximations and $\chi$SR phase (blue dashed).}\label{mp_S}
	\end{center}
\end{figure}
\begin{figure}[!htb]
	\begin{center}
	\includegraphics[width=8.5cm]{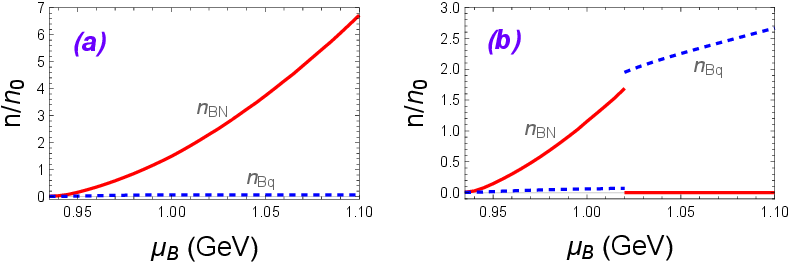}
		\caption{For the SQVM set of parameters with $(a)\ d=1$ and $(b)\ d=3/2$ for ${\tilde{G}_3}$, the contributions of nucleons (red solid) and quarks (blue dashed) to baryon density $n_{\rm B}$ in the ground states.}\label{nB_S}
	\end{center}
\end{figure}

At zero temperature, the numerical results are presented in Figs.~\ref{mp_S}-\ref{nB_S} for the SQVM set of parameters. In mean field approximation, the thermodynamic potential is given by Eq.\eqref{omg0}, and the blue dashed lines in Figs.\ref{mp_S} are the corresponding results for the $\chi$SR phase. In this approximation, the quark mass keeps around $m_{\rm q}=0.3135~{\rm GeV}$ for the $\chi$SB phase (not shown in the figures); so chiral restoration takes place through a first-order jump to a much smaller mass~\cite{Klevansky:1992qe,Hatsuda:1994pi}, $m_{\rm q}\approx0.1~{\rm GeV}$, see the left ends of the blue dashed lines in the upper panels of Fig.~\ref{mp_S}. Here and in the following, the pressure $p$ is related to the thermodynamic potential $\Omega$ as $p=\Omega_{\rm v}-\Omega$ with $\Omega_{\rm v}$ the vacuum part. 

Beyond mean field (BMF) approximation, the thermodynamic potential is given by Eq.\eqref{omg} with $\Omega_\pi=0$ at zero temperature. There are two different ways to determine the quark mass: One is to refer to the mean-field gap equation $\partial\Omega_0/\partial m=0$, and the Goldstone theorem is guaranteed; the other one is to find the minimum of Eq.\eqref{omg} by taking into account Eqs.\eqref{Nmass} self-consistently. We refer to the two ways as inconsistent beyond mean field (IBMF) and consistent beyond mean field (CBMF) approximations, respectively. The former way was adopted to study BCS-BEC crossover in a two dimensional system, and all the mean field results were improved to be more consistent with the Monte-Carlo simulations~\cite{He:2015nxa}. The corresponding pressures are compared in the lower panels of Figs.~\ref{mp_S}. Nevertheless, the results of BMF approximation are the same as those of MF approximation for the $\chi$SR phase thus can be presented by single lines, see the blue dashed lines. That is because $m_{\rm N}>\mu_{\rm B}$ or no solution can be found from Eq.\eqref{Nmass} for a small $m_{\rm q}$, and then $\Omega_{\rm N}=0$ at zero temperature. In this sense, chiral symmetry restoration transition, if occurs, is also deconfinement transition, so we would generally use "phase transition" in the following.

In the IBMF approximation, the pressures of $\chi$SB phase become smaller than those of $\chi$SR phase beyond the intersections, so phase transition happens and is of first order for both $d$. However, the decreasing of pressures with $\mu_{\rm B}$ just before the transition points would indicate negative baryon densities thus is not physical. In the CBMF approximation for $d=1$, the pressure of $\chi$SB phase is always larger than that of $\chi$SR phase, so no phase transition occurs at all. While in the CBMF approximation for $d=3/2$, the pressure of $\chi$SB phase is exceeded by that of $\chi$SR phase at $\mu_{\rm B}=1.02~{\rm GeV}$ and the phase transition is of first order. In fact, even though the quark mass is almost the same, the results are quite subtle to the effective nucleon mass: $m_{\rm N}$ is a bit larger in the $\chi$SB phase for $d=3/2$ thus the pressure is suppressed more, compare the upper panels of Fig.~\ref{mp_S}.  { One might notice that the quark mass first slightly decrease and then increases with $\mu_{\rm B}$ in the chiral symmetry breaking phase. To my understanding, this is a competing result between the $\mu_{\rm B}$ effects from quarks and nucleons: $\mu_{\rm B}$ tends to reduce quark dynamical mass according to chiral symmetry restoration; but to maintain stable nucleons with a small mass, it prefers the quarks to stay in the chiral symmetry breaking (and also confined) phase.} The corresponding baryon densities are illustrated in Fig.~\ref{nB_S} for the ground states. It is clear that nucleons overwhelmingly dominate the $\chi$SB phase thus give rise to confined nuclear matter, while quarks overwhelmingly dominate the $\chi$SR phase thus give rise to deconfined quark matter. Usually, we expect the basic degrees of freedom to change from nucleons to quarks with increasing baryon density, so the results with $d=3/2$ is more reasonable. In this case, the total baryon density weakly jumps up from nuclear matter to quark matter around $n_{\rm B}=2n_0$, and no quarkyonic matter is found to exist here. {Though, it is possible that nucleons exist as unbound resonances in quark matter when applying a ring sum of nucleonic polarizations.}

 \subsection{Large quark vacuum mass}
 \begin{figure}[!htb]
	\begin{center}
	\includegraphics[width=8.5cm]{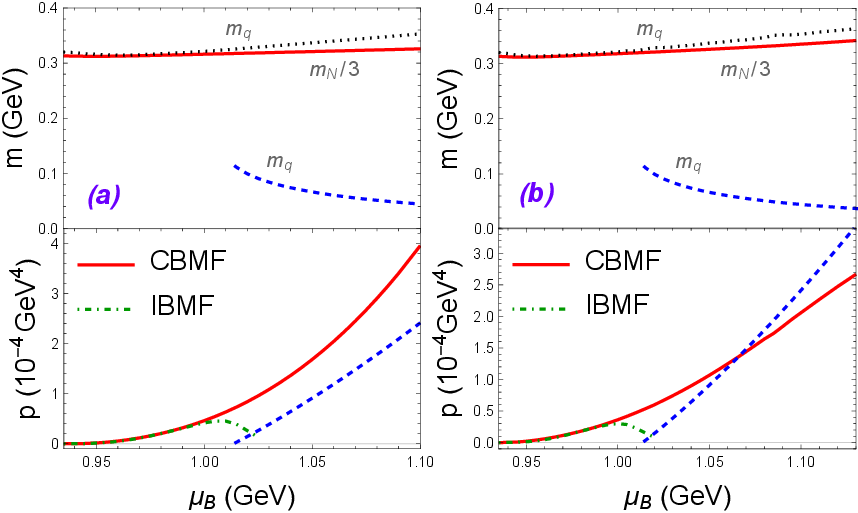}
		\caption{The LQVM case with $(a)\ d=1$ and $(b)\ d=3/2$ for ${\tilde{G}_3}$. The notations follow Fig.\ref{mp_S}.}\label{mp_L}
	\end{center}
\end{figure}
\begin{figure}[!htb]
	\begin{center}
	\includegraphics[width=8.5cm]{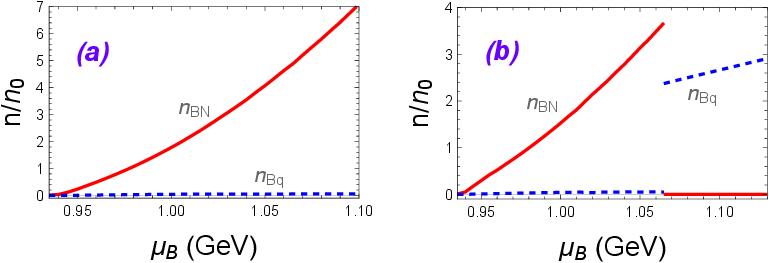}
		\caption{For the LQVM set of parameters with $(a)\ d=1$ and $(b)\ d=3/2$ for ${\tilde{G}_3}$, the contributions of nucleons (red solid) and quarks (blue dashed) to baryon density $n_{\rm B}$ in the ground states.}\label{nB_L}
	\end{center}
\end{figure}
At zero temperature, the numerical results are presented in Figs.~\ref{mp_L}-\ref{nB_L}  for the LQVM set of parameters.  For $d=1$, all the features are similar to those of the SQVM case. For $d=3/2$, there is also a first order phase transition from nuclear matter to quark matter around $\mu_{\rm B}=1.065~{\rm GeV}$ according to the lower panel of Fig.~\ref{mp_L}$(b)$. However, the critical chemical potential is so large that a larger baryon density would be established in the nuclear matter than in the quark matter at the transition point, see the right panel of Fig.~\ref{nB_L}. That is weird, maybe saturation is the key point to cure the problem since the pressure, thus the baryon density, would be smaller in the nuclear matter phase. 

\section{Summary}\label{summary}
In this work, we extend the two-flavor Nambu--Jona-Lasinio model to include a chiral symmetric six-quark interaction term, standing for three quarks interacting via nucleonic channel. Then, similar to mesons, nucleons could emerge as collective excitations of quarks with their masses evaluated in random phase approximation. Beyond mean field approximation, collective excitations also contribute to the thermodynamic potential and are simply presented in pole approximation here. For all the cases considered, quarkyonic matter is never found to exist due to the great enhancement of nucleon masses with chiral symmetry restoration, thus only nuclear and quark matters are possible phases of cold, dense QCD matter. For $d=1$, chiral symmetry never restores in the consistent scheme and nuclear matter would survive to a very large baryon density; while for $d=3/2$, chiral symmetry gets restored through a first-order transition and the QCD system becomes quark matter at a large baryon density. It is well accepted that the dominant degrees of freedom would alter from nucleons to quarks with increasing baryon density, since the mean distance between nucleons would become smaller than the diameter of nucleons. In this sense, only the case with stronger confinement effect and a smaller quark mass gives reasonable results, and the corresponding first-order transition occurs weakly around the baryon density $2n_0$.

In this model, quarks and nucleons can be treated at the same level, thus it is very suitable for a consistent and systematic study of transitions among nuclear, quark and quarkyonic matters. In the future, several important issues must be taken care of, such as accounting for feedback of nucleonic pairing to quarks and reproducing saturation property of nuclear matter. 

\acknowledgments
G.C. is supported by the Natural Science Foundation of Guangdong Province with Grant No. 2024A1515011225.
\appendix
\begin{widetext}
\section{Evaluation of $\Pi_{\rm q\Delta}(P)$}\label{APiqd}
According to Eq.\eqref{PiD}, we have
\bea
\Pi_\Delta^{\rm c}(Q)&=&4\int{\di^4 k\over (2\pi)^4} {\rm Tr} {1\over \slashed{Q}-\slashed{k}+\gamma^0{\mu_{\rm B}\over 3}-m}{1\over \slashed{k}+\gamma^0{\mu_{\rm B}\over3}-m}\nonumber\\
&=&16\int{\di^4k\over (2\pi)^4}{(Q_0-k_0+{\mu_{\rm B}\over 3})(k_0+{\mu_{\rm B}\over 3})-({\bf Q}-{\bf k})\cdot{\bf k}+m^2\over \left[(Q_0-k_0+{\mu_{\rm B}\over 3})^2-E_{{\bf Q}-{\bf k}}^2\right]\left[(k_0+{\mu_{\rm B}\over 3})^2-E_{\bf k}^2\right]}\nonumber\\
&=&-4i\sum_{t=\pm}\int^\Lambda{\di^3{\bf k}\over (2\pi)^3}\left\{{(Q_0-t\,E_{\bf k}+{2\mu_{\rm B}\over 3})t\,E_{\bf k}-({\bf Q}-{\bf k})\cdot{\bf k}+m^2\over E_{\bf k}\left[(Q_0-t\,E_{\bf k}+{2\mu_{\rm B}\over 3})^2-E_{{\bf Q}-{\bf k}}^2\right]}\tanh{E_{\bf k}-t{\mu_{\rm B}\over 3}\over 2T}\right.\nonumber\\
&&\left.+{(Q_0-t\,E_{{\bf Q}-{\bf k}}+{2\mu_{\rm B}\over 3})t\,E_{{\bf Q}-{\bf k}}-({\bf Q}-{\bf k})\cdot{\bf k}+m^2\over E_{{\bf Q}-{\bf k}}\left[(Q_0-t\,E_{{\bf Q}-{\bf k}}+{2\mu_{\rm B}\over 3})^2-E_{{\bf k}}^2\right]}\tanh{E_{{\bf Q}-{\bf k}}-t{\mu_{\rm B}\over 3}\over 2T}\right\}\nonumber\\
&=&-8i\sum_{t=\pm}\int^\Lambda{\di^3{\bf k}\over (2\pi)^3}{(Q_0-t\,E_{\bf k}+{2\mu_{\rm B}\over 3})t\,E_{\bf k}-({\bf Q}-{\bf k})\cdot{\bf k}+m^2\over E_{\bf k}\left[(Q_0-t\,E_{\bf k}+{2\mu_{\rm B}\over 3})^2-E_{{\bf Q}-{\bf k}}^2\right]}\tanh{E_{\bf k}-t{\mu_{\rm B}\over 3}\over 2T}.
\eea
In the last step, the variable transformation ${\bf Q}-{\bf k}\leftrightarrow {\bf k}$ has been performed to the second term in the integrand, which then becomes the same as the first term. Here and in the following, to simplify the expressions, we assume that the cutoff $\Lambda$ to the three-momentum integral is large enough to ensure the validity of such a transformation. 

Then, remembering the summation over the color index $c$ in Eq.\eqref{PiqD} for $\Pi_{\rm q\Delta}(P)$, we have
\bea
&&\Pi_{\rm q\Delta}(P)={\tilde{G}_3}\int{\di^4 l\over (2\pi)^4}S_{\rm q}^{\rm c}(P-l)\Pi_\Delta^{\rm c}(l)=3i\,{\tilde{G}_3}\int{\di^4 l\over (2\pi)^4}{\gamma\cdot({ P-l})+\gamma^0{\mu_{\rm B}/3}+m\over (P_0-l_0+{\mu_{\rm B}\over 3})^2-E_{\bf P-l}^2}\Pi_\Delta^{\rm c}(l)\nonumber\\
&=&{3{\tilde{G}_3}\over4}\sum_{s=\pm}\int^\Lambda{\di^3 {\bf l}\over (2\pi)^3}\Bigg\{{s\, E_{\bf P-l}\gamma^0-\boldsymbol{\gamma}\cdot({\bf P-l})+m\over E_{\bf P-l}}\Pi_\Delta^{\rm c}(P_0+{\mu_{\rm B}\over 3}-s\, E_{\bf P-l},{\bf l})\tanh{E_{\bf P-l}-s{\mu_{\rm B}\over 3}\over 2T}-8i\sum_{t=\pm}\int^\Lambda{\di^3{\bf k}\over (2\pi)^3}\nonumber\\
&&{\gamma^0(P_0\!+\!\mu_{\rm B}\!-\!t\,E_{\bf k}\!-\!s\,E_{{\bf l}-{\bf k}})\!-\!\boldsymbol{\gamma}\cdot({\bf P\!-\!l})\!+\!m\over (P_0+{\mu_{\rm B}}-t\,E_{\bf k}-s\,E_{{\bf l}-{\bf k}})^2-E_{\bf P-l}^2}{s\,t\,E_{{\bf l}-{\bf k}}E_{\bf k}-({\bf l}-{\bf k})\cdot{\bf k}+m^2\over E_{\bf k}E_{{\bf l}-{\bf k}}}\tanh{E_{\bf k}-t{\mu_{\rm B}\over 3}\over 2T}\coth{E_{{\bf l}-{\bf k}}+s(t\,E_{\bf k}-{2\mu_{\rm B}\over 3})\over 2T}\Bigg\}\nonumber\\
&=&-{6i\,{\tilde{G}_3}}\sum_{s,t=\pm}\int^\Lambda{\di^3 {\bf l}\over (2\pi)^3}\int^\Lambda{\di^3{\bf k}\over (2\pi)^3}\tanh{E_{\bf k}-t{\mu_{\rm B}\over 3}\over 2T}\Bigg\{{s\, E_{\bf P-l}\gamma^0\!-\!\boldsymbol{\gamma}\cdot({\bf P\!-\!l})\!+\!m\over E_{\bf P-l}}{(\tilde{P}_0\!-\!t\,E_{\bf k}\!-\!s\, E_{\bf P-l})t\,E_{\bf k}\!-\!({\bf l}\!-\!{\bf k})\cdot{\bf k}\!+\!m^2\over E_{\bf k}\left[(\tilde{P}_0-t\,E_{\bf k}-s\, E_{\bf P-l})^2-E_{{\bf l}-{\bf k}}^2\right]}\nonumber\\
&&\!\!\tanh{E_{\bf P-l}\!-\!s{\mu_{\rm B}\over 3}\over 2T}+{\gamma^0(\tilde{P}_0\!-\!t\,E_{\bf k}\!-\!s\,E_{{\bf l}-{\bf k}})\!-\!\boldsymbol{\gamma}\cdot({\bf P\!-\!l})\!+\!m\over (\tilde{P}_0-t\,E_{\bf k}-s\,E_{{\bf l}-{\bf k}})^2-E_{\bf P-l}^2}{s\,t\,E_{{\bf l}-{\bf k}}E_{\bf k}\!-\!({\bf l}\!-\!{\bf k})\cdot{\bf k}\!+\!m^2\over E_{\bf k}E_{{\bf l}-{\bf k}}}\coth{E_{{\bf l}-{\bf k}}\!+\!s(t\,E_{\bf k}\!-\!{2\mu_{\rm B}\over 3})\over 2T}\Bigg\}.
\eea
 For ${\bf P}=0$, the expression can be simplified as
\bea
&&\Pi_{\rm q\Delta}(\tilde{P}_0)=-{6i\,{\tilde{G}_3}}\sum_{s,t=\pm}\int^\Lambda{\di^3 {\bf l}\over (2\pi)^3}\int^\Lambda{\di^3{\bf k}\over (2\pi)^3}\tanh{E_{\bf k}-t{\mu_{\rm B}\over 3}\over 2T}\Bigg\{{s\, E_{\bf l}\gamma^0+m\over E_{\bf l}}{(\tilde{P}_0-s\, E_{\bf l})t\,E_{\bf k}-{\bf l\cdot k}\over E_{\bf k}\left[(\tilde{P}_0-t\,E_{\bf k}-s\, E_{\bf l})^2-E_{{\bf l}-{\bf k}}^2\right]}\nonumber\\
&&\ \ \ \ \ \ \tanh{E_{\bf l}-s{\mu_{\rm B}\over 3}\over 2T}+{\gamma^0(\tilde{P}_0-t\,E_{\bf k}-s\,E_{{\bf l}-{\bf k}})+m\over (\tilde{P}_0-t\,E_{\bf k}-s\,E_{{\bf l}-{\bf k}})^2-E_{\bf l}^2}{s\,t\,E_{{\bf l}-{\bf k}}E_{\bf k}-({\bf l}-{\bf k})\cdot{\bf k}+m^2\over E_{\bf k}E_{{\bf l}-{\bf k}}}\coth{E_{{\bf l}-{\bf k}}+s(t\,E_{\bf k}-{2\mu_{\rm B}\over 3})\over 2T}\Bigg\}\nonumber\\
&=&-{6i\,{\tilde{G}_3}}\sum_{s,t=\pm}\int^\Lambda{\di^3 {\bf l}\over (2\pi)^3}\int^\Lambda{\di^3{\bf k}\over (2\pi)^3}{\tanh{E_{\bf k}-t{\mu_{\rm B}\over 3}\over 2T}\over E_{\bf l}E_{\bf k}\left[(\tilde{P}_0-t\,E_{\bf k}-s\, E_{\bf l})^2-E_{{\bf l}-{\bf k}}^2\right]}\Big\{({s\, E_{\bf l}\gamma^0+m})[{(\tilde{P}_0-s\, E_{\bf l})t\,E_{\bf k}-{\bf l\cdot k}}]\nonumber\\
&&\tanh{E_{\bf l}-s{\mu_{\rm B}\over 3}\over 2T}+[{\gamma^0(\tilde{P}_0-t\,E_{\bf k}-s\,E_{{\bf l}})+m}][{s\,t\,E_{{\bf l}}E_{\bf k}+{\bf l}\cdot{\bf k}+m^2}]\coth{E_{{\bf l}}+s(t\,E_{\bf k}-{2\mu_{\rm B}\over 3})\over 2T}\Big\}\nonumber\\
&=&-{6i\,{\tilde{G}_3}}\sum_{s,t=\pm}\int^\Lambda{\di^3 {\bf l}\over (2\pi)^3}\int^\Lambda{\di^3{\bf k}\over (2\pi)^3}{1\over E_{\bf l}E_{\bf k}\left[(\tilde{P}_0\!-\!t\,E_{\bf k}\!-\!s\, E_{\bf l})^2\!-\!E_{{\bf l}-{\bf k}}^2\right]}\Big\{({s\, E_{\bf l}\gamma^0\!+\!m})[{(\tilde{P}_0\!-\!s\, E_{\bf l})t\,E_{\bf k}\!-\!{\bf l\cdot k}}]H^{\rm s\,t}({\bf k},{\bf l})\nonumber\\
&&+{1\over2}[{\gamma^0(\tilde{P}_0-t\,E_{\bf k}-s\,E_{{\bf l}})+m}][{s\,t\,E_{{\bf l}}E_{\bf k}+{\bf l}\cdot{\bf k}+m^2}][s\,t\,+H^{\rm s\,t}({\bf k},{\bf l})]\Big\}\nonumber
\eea
with $H^{\rm s\,t}({\bf k},{\bf l})\equiv\tanh{E_{\bf k}-t{\mu_{\rm B}\over 3}\over 2T}\tanh{E_{{\bf l}}-s\,{\mu_{\rm B}\over 3}\over 2T}$. Here, regarding the second term in the integrand, we have transformed ${\bf l}-{\bf k}$ to $-{\bf l}$ in the second step and used symmetries under $s\leftrightarrow t,{\bf k}\leftrightarrow{\bf l}$ in the last step. To facilitate numerical calculations, we complete the integration over the angle between ${\bf l}$ and ${\bf k}$, and find
\bea
&&\!\!\!\!\!\!\!\!\Pi_{\rm q\Delta}(\tilde{P}_0)=-{6i\,{\tilde{G}_3}}\!\sum_{s,t=\pm}\!\int^\Lambda\!\!\!{l\,\di l\,k\,\di k\over 16\pi^4E_{\bf l}E_{\bf k}}L^{\rm s\,t}(\tilde{P}_0,k,l)\Bigg\{({s\, E_{\bf l}\gamma^0\!+\!m})\left[{(\tilde{P}_0\!-\!s\, E_{\bf l})t\,E_{\bf k}\!+\!{1\over2}\left((\tilde{P}_0\!-\!t\,E_{\bf k}\!-\!s\, E_{\bf l})^2\!-\!l^2\!-\!k^2\!-\!m^2\right)}\right]\nonumber\\
&&\ \ \ \ \ H^{\rm s\,t}({\bf k},{\bf l})+{1\over2}[{\gamma^0(\tilde{P}_0-t\,E_{\bf k}-s\,E_{{\bf l}})+m}]\left[{s\,t\,E_{{\bf l}}E_{\bf k}-{1\over2}\left((\tilde{P}_0\!-\!t\,E_{\bf k}\!-\!s\, E_{\bf l})^2\!-\!l^2\!-\!k^2\!-\!3m^2\right)}\right][s\,t\,+H^{\rm s\,t}({\bf k},{\bf l})]\Bigg\}\nonumber\\
&&\ \ \ \ \ -{6i\,{\tilde{G}_3}}\sum_{s,t=\pm}\int^\Lambda{l^2\,\di l\,k^2\,\di k\over 8\pi^4E_{\bf l}E_{\bf k}}\Big\{-({s\, E_{\bf l}\gamma^0+m})H^{\rm s\,t}({\bf k},{\bf l}) +{1\over2}[{\gamma^0(\tilde{P}_0-t\,E_{\bf k}-s\,E_{{\bf l}})+m}][s\,t\,+H^{\rm s\,t}({\bf k},{\bf l})]\Big\}\nonumber\\
&=&-{6i\,{\tilde{G}_3}}\!\sum_{s,t=\pm}\!\int^\Lambda\!\!\!{l\,\di l\,k\,\di k\over 16\pi^4E_{\bf l}E_{\bf k}}L^{\rm s\,t}(\tilde{P}_0,k,l)\Bigg\{({s\, E_{\bf l}\gamma^0\!+\!m})\left[{(\tilde{P}_0\!-\!s\, E_{\bf l})t\,E_{\bf k}\!+\!{1\over2}\left((\tilde{P}_0\!-\!t\,E_{\bf k}\!-\!s\, E_{\bf l})^2\!-\!l^2\!-\!k^2\!-\!m^2\right)}\right]\nonumber\\
&&\ \ \ \ \ H^{\rm s\,t}({\bf k},{\bf l})+{1\over2}[{\gamma^0(\tilde{P}_0-t\,E_{\bf k}-s\,E_{{\bf l}})+m}]\left[{s\,t\,E_{{\bf l}}E_{\bf k}-{1\over2}\left((\tilde{P}_0\!-\!t\,E_{\bf k}\!-\!s\, E_{\bf l})^2\!-\!l^2\!-\!k^2\!-\!3m^2\right)}\right][s\,t\,+H^{\rm s\,t}({\bf k},{\bf l})]\Bigg\}\nonumber\\
&&-{6i\,{\tilde{G}_3}}\sum_{s,t=\pm}\int^\Lambda{l^2\,\di l\,k^2\,\di k\over 8\pi^4E_{\bf l}E_{\bf k}}\left\{{1\over2}[{\gamma^0(\tilde{P}_0\!-\!t\,E_{\bf k}\!-\!s\,E_{{\bf l}})\!+\!m}]\!-\!({s\, E_{\bf l}\gamma^0\!+\!m})\right\}H^{\rm s\,t}({\bf k},{\bf l})\label{EPiqd}
\eea
with $L^{\rm s\,t}(\tilde{P}_0,k,l)\equiv\ln\left|{(\tilde{P}_0-t\,E_{\bf k}-s\, E_{\bf l})^2-E_{{l}-{k}}^2\over(\tilde{P}_0-t\,E_{\bf k}-s\, E_{\bf l})^2-E_{{l}+{k}}^2}\right|$. If we define $\Pi_{\rm q\Delta}(\tilde{P}_0)\equiv i\,\left[\gamma^0\Pi_{\rm q\Delta}^0(\tilde{P}_0)+\Pi_{\rm q\Delta}^{\rm s}(\tilde{P}_0)\right]$, then the time and scalar components are, respectively,
\bea
\Pi_{\rm q\Delta}^0(\tilde{P}_0)&=&-{6{\tilde{G}_3}}\sum_{s,t=\pm}\int^\Lambda{l\,\di l\,k\,\di k\over 16\pi^4E_{\bf l}E_{\bf k}}L^{\rm s\,t}(\tilde{P}_0,k,l)\Bigg\{{s\, E_{\bf l}}\left[{(\tilde{P}_0-s\, E_{\bf l})t\,E_{\bf k}+{1\over2}\left((\tilde{P}_0\!-\!t\,E_{\bf k}\!-\!s\, E_{\bf l})^2\!-\!l^2\!-\!k^2\!-\!m^2\right)}\right]\nonumber\\
&&H^{\rm s\,t}({\bf k},{\bf l})+{1\over2}{(\tilde{P}_0-t\,E_{\bf k}-s\,E_{{\bf l}})}\left[{s\,t\,E_{{\bf l}}E_{\bf k}-{1\over2}\left((\tilde{P}_0\!-\!t\,E_{\bf k}\!-\!s\, E_{\bf l})^2\!-\!l^2\!-\!k^2\!-\!3m^2\right)}\right][s\,t\,+H^{\rm s\,t}({\bf k},{\bf l})]\Bigg\}\nonumber\\
&&-{3{\tilde{G}_3}}\sum_{s,t=\pm}\int^\Lambda{l^2\,\di l\,k^2\,\di k\over 8\pi^4E_{\bf l}E_{\bf k}}{(\tilde{P}_0-4t\,E_{\bf k})}H^{\rm s\,t}({\bf k},{\bf l}),\\
\Pi_{\rm q\Delta}^{\rm s}(\tilde{P}_0)&=&-{6m{\tilde{G}_3}}\!\sum_{s,t=\pm}\!\int^\Lambda\!\!\!{l\,\di l\,k\,\di k\over 16\pi^4E_{\bf l}E_{\bf k}}L^{\rm s\,t}(\tilde{P}_0,k,l)\Bigg\{\left[{(\tilde{P}_0-s\, E_{\bf l})t\,E_{\bf k}+{1\over2}\left((\tilde{P}_0\!-\!t\,E_{\bf k}\!-\!s\, E_{\bf l})^2\!-\!l^2\!-\!k^2\!-\!m^2\right)}\right]H^{\rm s\,t}({\bf k},{\bf l})\nonumber\\
&&\!\!\!\!\!\!\!\!\!\!\!\!\!\!+{1\over2}\left[{s\,t\,E_{{\bf l}}E_{\bf k}\!-\!{1\over2}\left((\tilde{P}_0\!-\!t\,E_{\bf k}\!-\!s\, E_{\bf l})^2\!-\!l^2\!-\!k^2\!-\!3m^2\right)}\right][s\,t\,\!+\!H^{\rm s\,t}({\bf k},{\bf l})]\Bigg\}\!+\!{3m{\tilde{G}_3}}\!\sum_{s,t=\pm}\!\int^\Lambda\!\!{l^2\,\di l\,k^2\,\di k\over 8\pi^4E_{\bf l}E_{\bf k}}H^{\rm s\,t}({\bf k},{\bf l}).\label{Piqds}
\eea
In the vacuum with $T=\mu_{\rm B}=0$, they are reduced to
\bea
\Pi_{\rm q\Delta}^{0}(P)&=&-{6{\tilde{G}_3}}\sum_{s,t=\pm}\int^\Lambda{l\,\di l\,k\,\di k\over 16\pi^4E_{\bf l}E_{\bf k}}L^{\rm s\,t}(P_0,k,l)\Bigg\{{s\, E_{\bf l}}\left[{(P_0-s\, E_{\bf l})t\,E_{\bf k}+{1\over2}\left((P_0\!-\!t\,E_{\bf k}\!-\!s\, E_{\bf l})^2\!-\!l^2\!-\!k^2\!-\!m^2\right)}\right]\nonumber\\
&&\!\!\!\!\!\!\!\!\!\!\!+{(P_0\!-\!t\,E_{\bf k}\!-\!s\,E_{{\bf l}})}\left[{s\,t\,E_{{\bf l}}E_{\bf k}\!-\!{1\over2}\left((P_0\!-\!t\,E_{\bf k}\!-\!s\, E_{\bf l})^2\!-\!l^2\!-\!k^2\!-\!3m^2\right)}\right]\delta_{s,t}\Bigg\}-{3P_0{\tilde{G}_3}}\sum_{s,t=\pm}\int^\Lambda\!{l^2\,\di l\,k^2\,\di k\over 8\pi^4E_{\bf l}E_{\bf k}},\\
\Pi_{\rm q\Delta}^{\rm s}(P_0)&=&-{6m{\tilde{G}_3}}\sum_{s,t=\pm}\int^\Lambda{l\,\di l\,k\,\di k\over 16\pi^4E_{\bf l}E_{\bf k}}L^{\rm s\,t}(P_0,k,l)\Bigg\{{(P_0-s\, E_{\bf l})t\,E_{\bf k}+{1\over2}\left((P_0\!-\!t\,E_{\bf k}\!-\!s\, E_{\bf l})^2\!-\!l^2\!-\!k^2\!-\!m^2\right)}\nonumber\\
&&+\left[{s\,t\,E_{{\bf l}}E_{\bf k}-{1\over2}\left((P_0\!-\!t\,E_{\bf k}\!-\!s\, E_{\bf l})^2\!-\!l^2\!-\!k^2\!-\!3m^2\right)}\right]\delta_{s,t}\Bigg\}+{3m{\tilde{G}_3}}\sum_{s,t=\pm}\int^\Lambda{l^2\,\di l\,k^2\,\di k\over 8\pi^4E_{\bf l}E_{\bf k}}.
\eea

\section{Evaluation of $\Pi_{\rm qqq}(P)$}\label{APiqqq}
According to Eq.\eqref{Piqqq}, we have
\bea
\tilde\Pi_{\rm qqq}(P,l,k)&=&-6S_{\rm q}(P-l)S_{\rm q}(l-k)S_{\rm q}(k)=6i{\gamma\cdot({ P\!-\!l})\!+\!\gamma^0{\mu_{\rm B}\over3}\!+\!m\over (P_0\!-\!l_0\!+\!{\mu_{\rm B}\over 3})^2\!-\!E_{\bf P-l}^2}{\gamma\cdot({ l\!-\!k})\!+\!\gamma^0{\mu_{\rm B}\over3}\!+\!m\over (l_0\!-\!k_0\!+\!{\mu_{\rm B}\over 3})^2\!-\!E_{\bf l-k}^2}{\gamma\cdot{k}\!+\!\gamma^0{\mu_{\rm B}\over3}\!+\!m\over (k_0\!+\!{\mu_{\rm B}\over 3})^2\!-\!E_{\bf k}^2}.
\eea
For ${\bf P}=0$, remembering the integrations over ${\bf k}$ and ${\bf l}$ in Eq.\eqref{Piqqq0}, the invariance of the denominator under the transformations ${\bf k}\rightarrow-{\bf k}$ and ${\bf l}\rightarrow-{\bf l}$ can help to reduce the effective self-energy to 
\bea
\tilde\Pi_{\rm qqq}(P,l,k)={6i \,{\cal{N}}\over [(P_0-l_0+{\mu_{\rm B}\over 3})^2-E_{\bf l}^2][(l_0-k_0+{\mu_{\rm B}\over 3})^2-E_{\bf l-k}^2][(k_0+{\mu_{\rm B}\over 3})^2-E_{\bf k}^2]}
\eea
with the numerator evaluated as
\bea
{\cal{N}}&\equiv&[\gamma^0(P_0-l_0+{\mu_{\rm B}\over 3})+m][\gamma^0(l_0-k_0+{\mu_{\rm B}\over 3})+m][\gamma^0(k_0+{\mu_{\rm B}\over 3})+m]+{\bf l\cdot(l-k)}[\gamma^0(k_0+{\mu_{\rm B}\over 3})+m]\nonumber\\
&&\ \ \ \ +{\bf l\cdot k}[-\gamma^0(l_0-k_0+{\mu_{\rm B}\over 3})+m]-{\bf k\cdot(l-k)}[\gamma^0(P_0-l_0+{\mu_{\rm B}\over 3})+m]\nonumber\\
&=&\gamma^0\left[(P_0-l_0+{\mu_{\rm B}\over 3})(l_0-k_0+{\mu_{\rm B}\over 3})(k_0+{\mu_{\rm B}\over 3})+m^2\tilde{P}_0\right]+m\left[(P_0-l_0+{\mu_{\rm B}\over 3})(l_0-k_0+{\mu_{\rm B}\over 3})+\right.\nonumber\\
&&\ \ \ \ \left.(P_0-k_0+{2\mu_{\rm B}\over 3})(k_0+{\mu_{\rm B}\over 3})+m^2\right]+{\bf l\cdot(l-k)}[\gamma^0(k_0+{\mu_{\rm B}\over 3})+m]+{\bf l\cdot k}[-\gamma^0(l_0-k_0+{\mu_{\rm B}\over 3})+m]\nonumber\\
&&\ \ \ \ -{\bf k\cdot(l-k)}[\gamma^0(P_0-l_0+{\mu_{\rm B}\over 3})+m]\nonumber\\
&\stackrel{k\leftrightarrow l-k}=&\gamma^0\left[(P_0-l_0+{\mu_{\rm B}\over 3})(l_0-k_0+{\mu_{\rm B}\over 3})(k_0+{\mu_{\rm B}\over 3})+m^2\tilde{P}_0\right]+m\left[(2P_0-l_0-k_0+{\mu_{\rm B}})(k_0+{\mu_{\rm B}\over 3})+m^2\right]\nonumber\\
&&+2m\,{\bf l\cdot k}-{\bf k\cdot(l-k)}[\gamma^0(P_0-l_0+{\mu_{\rm B}\over 3})+m]\nonumber\\
&\stackrel{l\leftrightarrow P-l+k}=&\gamma^0\left[(P_0-l_0+{\mu_{\rm B}\over 3})(l_0-k_0+{\mu_{\rm B}\over 3})(k_0+{\mu_{\rm B}\over 3})+m^2\tilde{P}_0\right]+m\left[3(l_0-k_0+{\mu_{\rm B}\over3})(k_0+{\mu_{\rm B}\over 3})+m^2\right]\nonumber\\
&&+{\bf l\cdot k}[\gamma^0(l_0-k_0+{\mu_{\rm B}\over 3})+3m].\label{EPiqqq}
\eea
In the third and last steps, the notations $k\leftrightarrow l-k$ and $l\leftrightarrow P-l+k$ indicate that we have utilized the symmetries of four momentum integrals under these transformations. 
So, if we define $\Pi_{\rm qqq}(P)\equiv i\,\left[\gamma_0\Pi_{\rm qqq}^0(P)+\Pi_{\rm qqq}^{\rm s}(P)\right]$, the time and scalar components are in the following forms:
\bea
\Pi_{\rm qqq}^0(P)&=&6{\tilde{G}_3}\int{\di^4 l\over (2\pi)^4}{\di^4 k\over (2\pi)^4}{(P_0-l_0+{\mu_{\rm B}\over 3})(l_0-k_0+{\mu_{\rm B}\over 3})(k_0+{\mu_{\rm B}\over 3})+m^2\tilde{P}_0+{\bf l\cdot k}(l_0-k_0+{\mu_{\rm B}\over 3})\over [(P_0-l_0+{\mu_{\rm B}\over 3})^2-E_{\bf l}^2][(l_0-k_0+{\mu_{\rm B}\over 3})^2-E_{\bf l-k}^2][(k_0+{\mu_{\rm B}\over 3})^2-E_{\bf k}^2]},\\
\Pi_{\rm qqq}^{\rm s}(P)&=&6{\tilde{G}_3}m\int{\di^4 l\over (2\pi)^4}{\di^4 k\over (2\pi)^4}{3(l_0-k_0+{\mu_{\rm B}\over3})(k_0+{\mu_{\rm B}\over 3})+m^2+3{\bf l\cdot k}\over [(P_0-l_0+{\mu_{\rm B}\over 3})^2-E_{\bf l}^2][(l_0-k_0+{\mu_{\rm B}\over 3})^2-E_{\bf l-k}^2][(k_0+{\mu_{\rm B}\over 3})^2-E_{\bf k}^2]}.
\eea

Their explicit forms can be evaluated as the followings.
\bea
&&\!\!\!\!\!\!\!\!\!\!\!\!\!\!\!\!\!\!\!\!\Pi_{\rm qqq}^0(P)
=-i{3{\tilde{G}_3}\over2}\sum_{t=\pm}\int^\Lambda{\di^4 l\over (2\pi)^4}\int^\Lambda{\di^3 k\over (2\pi)^3}\left\{{(P_0-l_0+{\mu_{\rm B}\over 3})(l_0-t\,E_{\bf k}+{2\mu_{\rm B}\over 3})t\,E_{\bf k}+m^2\tilde{P}_0+{\bf l\cdot k}(l_0-t\,E_{\bf k}+{2\mu_{\rm B}\over 3})\over E_{\bf k}[(P_0-l_0+{\mu_{\rm B}\over 3})^2-E_{\bf l}^2][(l_0-t\,E_{\bf k}+{2\mu_{\rm B}\over 3})^2-E_{\bf l-k}^2]}\right.\nonumber\\
&&\left.\tanh{E_{\bf k}-t{\mu_{\rm B}\over 3}\over2T}+{(P_0-l_0+{\mu_{\rm B}\over 3})(l_0-t\,E_{\bf l-k}+{2\mu_{\rm B}\over 3})t\,E_{\bf l-k}+m^2\tilde{P}_0+t\,E_{\bf l-k}{\bf l\cdot k}\over E_{\bf l-k}[(P_0-l_0+{\mu_{\rm B}\over 3})^2-E_{\bf l}^2][(l_0-t\,E_{\bf l-k}+{2\mu_{\rm B}\over 3})^2-E_{\bf k}^2]}\tanh{E_{\bf l-k}-t{\mu_{\rm B}\over 3}\over2T}\right\}\nonumber\\
&\stackrel{\bf k\leftrightarrow l-k}=&\!\!\!\!-i{3{\tilde{G}_3}\over2}\sum_{t=\pm}\int^\Lambda\!\!\!{\di^4 l\over (2\pi)^4}\int^\Lambda\!\!\!{\di^3 k\over (2\pi)^3}{2(P_0\!-\!l_0\!+\!{\mu_{\rm B}\over 3})(l_0\!-\!t\,E_{\bf k}\!+\!{2\mu_{\rm B}\over 3})t\,E_{\bf k}\!+\!2m^2\tilde{P}_0\!+\!{\bf l\cdot k}(l_0\!-\!t\,E_{\bf k}\!+\!{2\mu_{\rm B}\over 3})\!+\!t\,E_{\bf k}{\bf l\cdot(l\!-\!k)}\over E_{\bf k}[(P_0-l_0+{\mu_{\rm B}\over 3})^2-E_{\bf l}^2][(l_0-t\,E_{\bf k}+{2\mu_{\rm B}\over 3})^2-E_{\bf l-k}^2]}\nonumber\\
&&\tanh{E_{\bf k}-t{\mu_{\rm B}\over 3}\over2T}\nonumber\\
&=&\!\!\!\!\!-{3{\tilde{G}_3}\over8}\!\!\sum_{t,s=\pm}\!\!\int^\Lambda\!\!\!\!{\di^3 l\over (2\pi)^3}\!\!\int^\Lambda\!\!\!\!{\di^3 k\over (2\pi)^3}\tanh{E_{\bf k}\!-\!t{\mu_{\rm B}\over 3}\over2T}\left\{{2(\tilde{P}_0\!-\!t\,E_{\bf k}\!-\!s\,E_{\bf l-k})s\,E_{\bf l-k}t\,E_{\bf k}\!+\!2m^2\tilde{P}_0\!+\!s\,E_{\bf l-k}{\bf l\cdot k}\!+\!t\,E_{\bf k}{\bf l\cdot(l\!-\!k)}\over E_{\bf k}E_{\bf l-k}[(\tilde{P}_0\!-\!t\,E_{\bf k}\!-\!s\,E_{\bf l-k})^2\!-\!E_{\bf l}^2]}\right.\nonumber\\
&&\!\!\!\!\!\left.\coth{E_{\bf l-k}+s\left(t\,E_{\bf k}-{2\mu_{\rm B}\over 3}\right)\over2T}+{(\tilde{P}_0-t\,E_{\bf k}-s\,E_{\bf l})(2s\,E_{\bf l}t\,E_{\bf k}+{\bf l\cdot k})+2m^2\tilde{P}_0+t\,E_{\bf k}{\bf l\cdot(l-k)}\over E_{\bf k}E_{\bf l}[(\tilde{P}_0-t\,E_{\bf k}-s\,E_{\bf l})^2-E_{\bf l-k}^2]}\tanh{E_{\bf l}-s{\mu_{\rm B}\over 3}\over2T}\right\}\nonumber\\
&\stackrel{\bf l\leftrightarrow k-l}=&\!\!\!\!-{3{\tilde{G}_3}\over8}\sum_{t,s=\pm}\int^\Lambda{\di^3 l\over (2\pi)^3}\int^\Lambda{\di^3 k\over (2\pi)^3}{\tanh{E_{\bf k}-t{\mu_{\rm B}\over 3}\over2T}\over E_{\bf k}E_{\bf l}[(\tilde{P}_0-t\,E_{\bf k}-s\,E_{\bf l})^2-E_{\bf l-k}^2]}\Big\{\Big[2(\tilde{P}_0-t\,E_{\bf k}-s\,E_{\bf l})s\,E_{\bf l}t\,E_{\bf k}+2m^2\tilde{P}_0\nonumber\\
&&\!\!\!\!+s\,E_{\bf l}{\bf (k-l)\cdot k}+t\,E_{\bf k}{\bf l\cdot(l-k)}\Big]\coth{E_{\bf l}+s\left(t\,E_{\bf k}-{2\mu_{\rm B}\over 3}\right)\over2T}+\Big[(\tilde{P}_0-t\,E_{\bf k}-s\,E_{\bf l})(2s\,E_{\bf l}t\,E_{\bf k}+{\bf l\cdot k})+2m^2\tilde{P}_0\nonumber\\
&&\!\!\!\!+t\,E_{\bf k}{\bf l\cdot(l-k)}\Big]\tanh{E_{\bf l}-s{\mu_{\rm B}\over 3}\over2T}\Big\}\nonumber\\
&\stackrel{s\leftrightarrow t,{\bf k}\leftrightarrow{\bf l}}=&-{3{\tilde{G}_3}\over8}\sum_{t,s=\pm}\int^\Lambda{\di^3 l\over (2\pi)^3}\int^\Lambda{\di^3 k\over (2\pi)^3}{1\over E_{\bf k}E_{\bf l}[(\tilde{P}_0-t\,E_{\bf k}-s\,E_{\bf l})^2-E_{\bf l-k}^2]}\Big\{\Big[(\tilde{P}_0-t\,E_{\bf k}-s\,E_{\bf l})s\,E_{\bf l}t\,E_{\bf k}+m^2\tilde{P}_0\nonumber\\
&&+t\,E_{\bf k}{\bf l\cdot(l-k)}\Big]\left[st\!+\!H^{\rm s\,t}({\bf k},{\bf l})\right]+\left[(\tilde{P}_0\!-\!t\,E_{\bf k}\!-\!s\,E_{\bf l})(2s\,E_{\bf l}t\,E_{\bf k}\!+\!{\bf l\cdot k})\!+\!2m^2\tilde{P}_0\!+\!t\,E_{\bf k}{\bf l\cdot(l\!-\!k)}\right]H^{\rm s\,t}({\bf k},{\bf l})\Big\}\nonumber\\
&=&-{3{\tilde{G}_3}\over8}\!\sum_{t,s=\pm}\!\int^\Lambda\!\!\!{l\,\di l\,k\,\di k\over 16\pi^4E_{\bf l}E_{\bf k}}L^{\rm s\,t}(\tilde{P}_0,k,l)\Big\{\Big[(\tilde{P}_0\!-\!t\,E_{\bf k}\!-\!s\,E_{\bf l})s\,E_{\bf l}t\,E_{\bf k}\!+\!m^2\tilde{P}_0\!+\!{1\over2}t\,E_{\bf k}((\tilde{P}_0\!-\!t\,E_{\bf k}\!-\!s\, E_{\bf l})^2\!+\!l^2\nonumber\\
&&-k^2\!-\!m^2)\Big]\left[st+H^{\rm s\,t}({\bf k},{\bf l})\right]+\left[(\tilde{P}_0-t\,E_{\bf k}-s\,E_{\bf l})\left(2s\,E_{\bf l}t\,E_{\bf k}-{1\over2}((\tilde{P}_0\!-\!t\,E_{\bf k}\!-\!s\, E_{\bf l})^2\!-\!l^2\!-\!k^2\!-\!m^2)\right)+2m^2\tilde{P}_0\right.\nonumber\\
&&\left.+{1\over2}t\,E_{\bf k}((\tilde{P}_0\!-\!t\,E_{\bf k}\!-\!s\, E_{\bf l})^2\!+\!l^2\!-\!k^2\!-\!m^2)\right]H^{\rm s\,t}({\bf k},{\bf l})\Big\}-{3{\tilde{G}_3}\over8}\!\sum_{s,t=\pm}\!\int^\Lambda\!\!{l^2\,\di l\,k^2\,\di k\over 8\pi^4E_{\bf l}E_{\bf k}}(\tilde{P}_0\!-\!4t\,E_{\bf k})H^{\rm s\,t}({\bf k},{\bf l}).
\eea
\bea
&&\!\!\!\!\!\!\!\!\!\!\!\!\!\!\!\!\!\!\!\!\Pi_{\rm qqq}^{\rm s}(P)=-i{3{\tilde{G}_3}\over2}m\sum_{t=\pm}\int{\di^4 l\over (2\pi)^4}\int^\Lambda{\di^3 k\over (2\pi)^3}\left\{{3(l_0-t\,E_{\bf k}+{2\mu_{\rm B}\over3})t\,E_{\bf k}+m^2+3{\bf l\cdot k}\over E_{\bf k}[(P_0\!-\!l_0\!+\!{\mu_{\rm B}\over 3})^2\!-\!E_{\bf l}^2][(l_0\!-\!t\,E_{\bf k}\!+\!{2\mu_{\rm B}\over 3})^2\!-\!E_{\bf l-k}^2]}\tanh{E_{\bf k}\!-\!t{\mu_{\rm B}\over 3}\over2T}\right.\nonumber\\
&&\left.+{3(l_0-t\,E_{\bf l-k}+{2\mu_{\rm B}\over3})t\,E_{\bf l-k}+m^2+3{\bf l\cdot k}\over E_{\bf l-k}[(P_0-l_0+{\mu_{\rm B}\over 3})^2-E_{\bf l}^2][(l_0-t\,E_{\bf l-k}+{2\mu_{\rm B}\over 3})^2-E_{\bf k}^2]}\tanh{E_{\bf l-k}-t{\mu_{\rm B}\over 3}\over2T}\right\}\nonumber\\
&\stackrel{\bf k\leftrightarrow l-k}=&-i{3{\tilde{G}_3}\over2}m\sum_{t=\pm}\int{\di^4 l\over (2\pi)^4}\int^\Lambda{\di^3 k\over (2\pi)^3}{2\left[3(l_0-t\,E_{\bf k}+{2\mu_{\rm B}\over3})t\,E_{\bf k}+m^2\right]+3{\bf l}^2\over E_{\bf k}[(P_0-l_0+{\mu_{\rm B}\over 3})^2-E_{\bf l}^2][(l_0-t\,E_{\bf k}+{2\mu_{\rm B}\over 3})^2-E_{\bf l-k}^2]}\tanh{E_{\bf k}-t{\mu_{\rm B}\over 3}\over2T}\nonumber\\
&=&-{3{\tilde{G}_3}\over8}m\sum_{t,s=\pm}\int^\Lambda{\di^3 l\over (2\pi)^3}\int^\Lambda{\di^3 k\over (2\pi)^3}\left\{{2\left[3s\,E_{\bf l-k}t\,E_{\bf k}+m^2\right]+3{\bf l}^2\over E_{\bf k}E_{\bf l-k}[(\tilde{P}_0-t\,E_{\bf k}-s\,E_{\bf l-k})^2-E_{\bf l}^2]}\coth{E_{\bf l-k}+s\left(t\,E_{\bf k}-{2\mu_{\rm B}\over 3}\right)\over2T}\right.\nonumber\\
&&\left.+{2\left[3(\tilde{P}_0-t\,E_{\bf k}-s\,E_{\bf l})t\,E_{\bf k}+m^2\right]+3{\bf l}^2\over E_{\bf k}E_{\bf l}[(\tilde{P}_0-t\,E_{\bf k}-s\,E_{\bf l})^2-E_{\bf l-k}^2]}\tanh{E_{\bf l}-s{\mu_{\rm B}\over 3}\over2T}\right\}\tanh{E_{\bf k}-t{\mu_{\rm B}\over 3}\over2T}\nonumber
\eea
\bea
&\stackrel{\bf l\leftrightarrow k-l}=&-{3{\tilde{G}_3}\over8}m\sum_{t,s=\pm}\int^\Lambda{\di^3 l\over (2\pi)^3}\int^\Lambda{\di^3 k\over (2\pi)^3}{\tanh{E_{\bf k}-t{\mu_{\rm B}\over 3}\over2T}\over E_{\bf k}E_{\bf l}[(\tilde{P}_0-t\,E_{\bf k}-s\,E_{\bf l})^2-E_{\bf l-k}^2]}\Big\{\left[2\left(3s\,E_{\bf l}t\,E_{\bf k}+m^2\right)+3({\bf l-k})^2\right]\nonumber\\
&&\coth{E_{\bf l}+s\left(t\,E_{\bf k}-{2\mu_{\rm B}\over 3}\right)\over2T}+\left[6(\tilde{P}_0-t\,E_{\bf k}-s\,E_{\bf l})t\,E_{\bf k}+2m^2+3{\bf l}^2\right]\tanh{E_{\bf l}-s{\mu_{\rm B}\over 3}\over2T}\Big\}\nonumber\\
&\stackrel{s\leftrightarrow t,{\bf k}\leftrightarrow{\bf l}}=&-{3{\tilde{G}_3}\over8}m\sum_{t,s=\pm}\int^\Lambda{\di^3 l\over (2\pi)^3}\int^\Lambda{\di^3 k\over (2\pi)^3}{1\over E_{\bf k}E_{\bf l}[(\tilde{P}_0-t\,E_{\bf k}-s\,E_{\bf l})^2-E_{\bf l-k}^2]}\Big\{\left[3s\,E_{\bf l}t\,E_{\bf k}+m^2+{3}{\bf l\cdot(l-k)}\right]\nonumber\\
&&\left[st+H^{\rm s\,t}({\bf k},{\bf l})\right]+\left[6(\tilde{P}_0-t\,E_{\bf k}-s\,E_{\bf l})t\,E_{\bf k}+2m^2+3{\bf l}^2\right]H^{\rm s\,t}({\bf k},{\bf l})\Big\}\nonumber\\
&=&-{3{\tilde{G}_3}\over8}m\sum_{t,s=\pm}\int^\Lambda{l\,\di l\,k\,\di k\over 16\pi^4E_{\bf l}E_{\bf k}}L^{\rm s\,t}(\tilde{P}_0,k,l)\Big\{\left[3s\,E_{\bf l}t\,E_{\bf k}+m^2+{3\over2}((\tilde{P}_0\!-\!t\,E_{\bf k}\!-\!s\, E_{\bf l})^2\!+\!l^2\!-\!k^2\!-\!m^2)\right]\nonumber\\
&&\!\!\!\!\!\!\left[st\!+\!H^{\rm s\,t}({\bf k},{\bf l})\right]+\left[6(\tilde{P}_0\!-\!t\,E_{\bf k}\!-\!s\,E_{\bf l})t\,E_{\bf k}\!+\!2m^2\!+\!3{\bf l}^2\right]H^{\rm s\,t}({\bf k},{\bf l})\Big\}+{9{\tilde{G}_3}\over8}m\!\sum_{s,t=\pm}\!\int^\Lambda\!\!{l^2\,\di l\,k^2\,\di k\over 8\pi^4E_{\bf l}E_{\bf k}}H^{\rm s\,t}({\bf k},{\bf l}).\label{Piqqqs}
\eea
In the vacuum with $T=\mu_{\rm B}=0$, they are reduced to
\bea
\Pi_{\rm qqq}^0(P)&=&-{3{\tilde{G}_3}\over8}\!\sum_{t,s=\pm}\!\int^\Lambda\!\!\!{l\,\di l\,k\,\di k\over 16\pi^4E_{\bf l}E_{\bf k}}L^{\rm s\,t}({P}_0,k,l)\Big\{\left[2({P}_0\!-\!t\,E_{\bf k}\!-\!s\,E_{\bf l})E_{\bf l}E_{\bf k}\!+\!2m^2{P}_0\!+\!t\,E_{\bf k}(({P}_0\!-\!t\,E_{\bf k}\!-\!s\, E_{\bf l})^2\!+\!l^2\right.\nonumber\\
&&\left.-k^2\!-\!m^2)\right]\delta_{\rm s,t}+\left[({P}_0-t\,E_{\bf k}-s\,E_{\bf l})(2s\,E_{\bf l}t\,E_{\bf k}-{1\over2}(({P}_0\!-\!t\,E_{\bf k}\!-\!s\, E_{\bf l})^2\!-\!l^2\!-\!k^2\!-\!m^2))+2m^2{P}_0\right.\nonumber\\
&&\left.+{1\over2}t\,E_{\bf k}(({P}_0\!-\!t\,E_{\bf k}\!-\!s\, E_{\bf l})^2\!+\!l^2\!-\!k^2\!-\!m^2)\right]\Big\}-{3{\tilde{G}_3}\over8}{P}_0\!\sum_{s,t=\pm}\!\int^\Lambda\!\!{l^2\,\di l\,k^2\,\di k\over 8\pi^4E_{\bf l}E_{\bf k}},\\
\Pi_{\rm qqq}^{\rm s}(P)&=&-{3{\tilde{G}_3}\over8}m\sum_{t,s=\pm}\int^\Lambda{l\,\di l\,k\,\di k\over 16\pi^4E_{\bf l}E_{\bf k}}L^{\rm s\,t}({P}_0,k,l)\Big\{\left[6E_{\bf l}E_{\bf k}+2m^2+{3}(({P}_0\!-\!t\,E_{\bf k}\!-\!s\, E_{\bf l})^2\!+\!l^2\!-\!k^2\!-\!m^2)\right]\delta_{\rm s,t}\nonumber\\
&&+\left[6(\tilde{P}_0\!-\!t\,E_{\bf k}\!-\!s\,E_{\bf l})t\,E_{\bf k}\!+\!2m^2\!+\!3{\bf l}^2\right]\Big\}+{9{\tilde{G}_3}\over8}m\sum_{s,t=\pm}\int^\Lambda{l^2\,\di l\,k^2\,\di k\over 8\pi^4E_{\bf l}E_{\bf k}}.
\eea
\end{widetext}


\begin{thebibliography}{99}
\bibitem{Nambu:1961tp}
Y.~Nambu and G.~Jona-Lasinio,
Phys. Rev. \textbf{122}, 345-358 (1961).

\bibitem{Nambu:1961fr}
Y.~Nambu and G.~Jona-Lasinio,
Phys. Rev. \textbf{124}, 246-254 (1961).

\bibitem{Gell-Mann:1964ewy}
M.~Gell-Mann,
Phys. Lett. \textbf{8}, 214-215 (1964).

\bibitem{Zweig:1964ruk}
G.~Zweig,
CERN-TH-401.

\bibitem{Klevansky:1992qe}
  S.~P.~Klevansky,
  The Nambu-Jona-Lasinio model of quantum chromodynamics,
  Rev.\ Mod.\ Phys.\  {\bf 64}, 649 (1992).

\bibitem{Hatsuda:1994pi} 
  T.~Hatsuda and T.~Kunihiro,
 QCD phenomenology based on a chiral effective Lagrangian,
  Phys.\ Rept.\  {\bf 247}, 221 (1994).

\bibitem{Luo:2017faz}
X.~Luo and N.~Xu,
Nucl. Sci. Tech. \textbf{28}, no.8, 112 (2017).

\bibitem{Luo:2022mtp}
X.~Luo, Q.~Wang, N.~Xu and P.~Zhuang,
Springer, 2022,
ISBN 978-981-19444-0-6, 978-981-19444-1-3.

\bibitem{He:2005nk}
L.~y.~He, M.~Jin and P.~f.~Zhuang,
Phys. Rev. D \textbf{71}, 116001 (2005).

\bibitem{Sun:2007fc}
G.~f.~Sun, L.~He and P.~Zhuang,
Phys. Rev. D \textbf{75}, 096004 (2007).

\bibitem{Cao:2016ats}
G.~Cao, L.~He and X.~G.~Huang,
Chin. Phys. C \textbf{41}, no.5, 051001 (2017).

\bibitem{Miransky:2015ava}
V.~A.~Miransky and I.~A.~Shovkovy,
Quantum field theory in a magnetic field: From quantum chromodynamics to graphene and Dirac semimetals,
Phys. Rept. \textbf{576}, 1-209 (2015).

\bibitem{Andersen:2014xxa}
J.~O.~Andersen, W.~R.~Naylor and A.~Tranberg,
Phase diagram of QCD in a magnetic field: A review,
Rev. Mod. Phys. \textbf{88}, 025001 (2016).

\bibitem{Cao:2021rwx}
G.~Cao,
Recent progresses on QCD phases in a strong magnetic field: views from Nambu\textendash{}Jona-Lasinio model,
Eur. Phys. J. A \textbf{57}, 264 (2021).

\bibitem{McLerran:2007qj}
L.~McLerran and R.~D.~Pisarski,
Nucl. Phys. A \textbf{796}, 83-100 (2007).

\bibitem{McLerran:2018hbz}
L.~McLerran and S.~Reddy,
Phys. Rev. Lett. \textbf{122}, no.12, 122701 (2019).

\bibitem{Cao:2020byn}
G.~Cao and J.~Liao,
JHEP \textbf{10}, 168 (2020).

\bibitem{Pisarski:2021aoz}
R.~D.~Pisarski,
Phys. Rev. D \textbf{103}, no.7, L071504 (2021).

\bibitem{Sen:2020qcd}
S.~Sen and L.~Sivertsen,
Astrophys. J. \textbf{915}, no.2, 109 (2021).

\bibitem{Cao:2022inx}
G.~Cao,
Phys. Rev. D \textbf{105}, no.11, 114020 (2022).

\bibitem{Koch:2022act}
V.~Koch and V.~Vovchenko,
Phys. Lett. B \textbf{841}, 137942 (2023).

\bibitem{Fujimoto:2023mzy}
Y.~Fujimoto, T.~Kojo and L.~D.~McLerran,
Phys. Rev. Lett. \textbf{132}, no.11, 112701 (2024).

\bibitem{Alford:2007xm}
M.~G.~Alford, A.~Schmitt, K.~Rajagopal and T.~Sch\"afer,
Rev. Mod. Phys. \textbf{80}, 1455-1515 (2008).

\bibitem{Buck:1992wz}
A.~Buck, R.~Alkofer and H.~Reinhardt,
Phys. Lett. B \textbf{286}, 29-35 (1992)

\bibitem{Ishii:1995bu}
N.~Ishii, W.~Bentz and K.~Yazaki,
Nucl. Phys. A \textbf{587}, 617-656 (1995).

\bibitem{Wang:2010iu}
J.~c.~Wang, Q.~Wang and D.~H.~Rischke,
Phys. Lett. B \textbf{704}, 347-353 (2011).

\bibitem{Ioffe:1981kw}
B.~L.~Ioffe,
Nucl. Phys. B \textbf{188}, 317-341 (1981).

\bibitem{Ioffe:1982ce}
B.~L.~Ioffe,
Z. Phys. C \textbf{18}, 67 (1983).

\bibitem{Jin:1993up}
X.~m.~Jin, M.~Nielsen, T.~D.~Cohen, R.~J.~Furnstahl and D.~K.~Griegel,
Phys. Rev. C \textbf{49}, 464-477 (1994).

\bibitem{Eichmann:2016yit}
G.~Eichmann, H.~Sanchis-Alepuz, R.~Williams, R.~Alkofer and C.~S.~Fischer,
Prog. Part. Nucl. Phys. \textbf{91}, 1-100 (2016).

\bibitem{He:2015nxa}
L.~He, H.~L\"u, G.~Cao, H.~Hu and X.~J.~Liu,
Phys. Rev. A \textbf{92}, no.2, 023620 (2015).

\bibitem{Zhuang:1994dw}
P.~Zhuang, J.~Hufner and S.~P.~Klevansky,
Nucl.\ Phys.\ A {\bf 576}, 525 (1994).

\end{thebibliography}
\end{document}